# Collective baryon decay and gravitational collapse

George Chapline[1] and James Barbieri[2]

[1]Lawrence Livermore National Laboratory, Livermore, CA
[2]NAWC-WD, China Lake, CA

While it is widely believed that the gravitational collapse of a sufficiently large mass will lead to a density singularity and an event horizon, we propose that this never happens when quantum effects are taken into account. In particular, we propose that when the conditions become ripe for a trapped surface to form, a quantum critical surface sweeps over the collapsing body, transforming the nucleons in the collapsing matter into a lepton/photon gas together with a positive vacuum energy. This will happen regardless of the matter density at the time a trapped surface starts to form, and as a result we predict that at least in all cases of gravitational collapse involving ordinary matter, a large fraction of the rest mass of the collapsing matter will be converted into a burst of neutrinos, and $\gamma$-rays. We predict that the peak luminosity of these bursts is only weakly dependent on the mass of the collapsing object, and is on the order of $(\varepsilon_q/m_P c^2)^{1/4} c^5/G$, where $\varepsilon_q$ is the mean energy of a nucleon parton and $m_P$ is the Planck mass. The duration of the bursts will depend the mass of the collapsing objects; in the case of stellar core collapse we predict that the duration of both the neutrino and $\gamma$-ray bursts will be on the order of 10 seconds.

The ultimate fate of matter undergoing gravitational collapse is a long standing enigma. Following the seminal paper of Oppenheimer and Snyder [1] it has come to be widely accepted that the gravitational collapse of a sufficiently large mass will inevitably lead to the formation of a density singularity and an event horizon [2]. Moreover, it has generally been believed that these classical predictions will turn out to be correct even when quantum effects are taken into account. This belief is based on the observation that if the collapsing mass is suffiently large, then the formation of an event horizon and initiation of irreversable collapse will take place in a region of space-time where the curvature of space is very small. Past experience suggests that classical general relativity ought to work well in such circumstances. On the other hand, there are number of long standing enigmas connected with the general relativisic picture of gravitational collapse. The most famous of these puzzles concerns the question of what happens to the quantum mechanical information carried by the collapsing matter; in quantum mechanics information can never disappear. The most likely resolution of this paradox is that that quantum effects can affect classical solutions of the Einstein equations in a singular way [3,4]. Indeed there are plausible arguments that in a quantum theory of emergent space-time, and irrespective of the local space-time curvature, quantum fields near to an event horizon will exhibit large fluctuations with a spectral bandwidth extending up to the Planck scale [5]. In previous papers [6,7] we have pointed out that as a consequence of these high frequency quantum fluctuations nucleons can be expected to decay as they approach a compact object which lies inside its gravitational radius. In fact there is circumstancial evidence, based on the observed spectrum of γ-rays coming from a "bubble" in the central region of our galaxy possibly associated with SgA* that protons falling onto the surface of a compact object do in fact produce positrons and γ-rays with the energies expected for nucleon decay [8].

In this paper we wish to draw attention to the possibility that essentially the same physics comes into play when a mass larger than the maximum mass of a neutron star undergoes gravitational collapse, and that as a consequence the final state of a body undergoing gravitational collapse will be dramatically different from that predicted by the classical general relativistic equations for gravitational collapse. In particular, we will argue that if the nucleons in the collapsing matter collectively decay due to quantum fluctuations associated with the tendency to form a trapped surface from which photons can not reach infinity, then neither a density singularity nor an event horizon will ever form.

Although it is uncertain when a trapped surface first appears in realistic examples of gravitational collapse, analytical studies suggest [9] that in general a trapped surface begins to form at the center of the collapsing object a short time before the outer radius of the collapsing mass reaches the gravitational radius. Classical general relativity predicts that this trapped surface will rapidly expand until it reaches the surface of the collapsing object and then relax by gravitational radiation to a stationary event horizon. Our interest is in the behavior of elementary particles in the interior of the collapsing mass as conditions become ripe for a trapped surface to form. We believe that for the same reasons that quantum fluctuations become large near to an event horizon, trapped surfaces are associated with large quantum fluctuations with a frequency spectrum extending up to the Planck frequency. As a result of these very high frequency fluctuations we expect that all the nucleons in any kind of matter undergoing gravitational collapse will collectively decay into a mixture of leptons, photons, and vacuum energy droplets. In particular, we argue that this is almost certainly what happens during the gravitational collapse of ordinary matter in "grand unified" (GUT) models for elementary particle interactions with baryon number violating gauge field couplings, such as the Georgi-Glashow model [10]. It is even conceivable that due to GUT symmery restoration at Planck scale energies that the nucleons in ordinary matter will decay coherently in a fashion resembling the superradiance that has recently been observed for nuclear x-ray transitions [11] .

In order to illustrate the effect of collective nucleon decay on the dynamics of collapse, we will reconsider the problem studied by Oppenheimer and Snyder: the gravitational collapse of a spherical homogeneous cloud of dust. We will assume that initial mass $M$ of the dust cloud is sufficiently large and dispersed so that up to the point where the cloud radius $R$ approaches its gravitational radius $R_g = 2GM/c^2$ the pressure is negligible. However, in a departure from Oppenheimer and Snyder's assumption that a collapsing dust cloud will remain homogeneous and pressure free right up until the time when a density singularity is formed, we will argue that in reality as the radius of a dust cloud approaches its Schwarzchild radius the dust cloud will be converted into a plasma with a zero average net baryon number density that is confined to a region of space-time with a large vacuum energy. In acordance with the views put forward in [3-5] as to how quantum mechanics changes the classical picture of compact objects, the region where the vacuum energy is large will be bounded by a quantum critical layer whose radius coincides with the classical event horizon radius $R_g$ and whose thickness can be estimated as [5,6]:

$$\Delta z = 2 R_g \left( \frac{\varepsilon_q}{m_P c^2} \right)^{1/2}, \qquad (1)$$

where $\varepsilon_q$ is the typical energy of a quark or gluon inside a nucleon, and $m_P \equiv (hc/G)^{1/2}$ is the Planck mass. After its germination near to the center of the collapsing dust cloud the quantum critical layer will expand, transforming the nucleons in the dust cloud into a gas of leptons and photons together with droplets of vacuum energy. The quantum crtical layer will expand in exactly the same way as the trapped surface that is predicted by classical general relativity. As is the case with the general relativistic trapped surface, our quantum critical layer will expand until it reaches the gravitational radius $R_g$' of the dust cloud plus nucleon decay products that have not yet escaped (we show below that a significant fraction of the rest mass-energy of the original dust will be radiated away by neutrinos by the time quantum critical layer reaches $R_g$').

Droplets of vacuum energy arise from collective baryon decay because the vacuum state in an asymptotically free gauge theory such as QCD is profoundly different from a Higgs vacuum with broken gauge symmetry. Futhermore if supersymmetry is broken, then these different kinds of vacuum states will have different energies. It follows that if both GUT symmetry and supersymmetry are broken, then when particles in the strongly confined sector of the GUT theory, viz. quarks and gluons, are suddenly transformed into particles that are not confined, viz. leptons and photons, then a vacuum energy will be left over. This vacuum energy represents the difference in the ground state energy between a gas of quarks and a gas of leptons with the same fermion number density. If the gauge field coupling constant in the confined sector much larger than the gauge field coulpling constant in the Higgs sector the difference in ground state energy density can be approximated as the exchange energy density of a uniform quark gas.

Of course if the density of nucleons as they approach the critical surface is smaller than the density of nucleons in nuclear matter, then the quarks would not have a uniform density. In such a case one might expect that collective decay of nucleons will initially give rise to droplets of vacuum energy rather than a uniform density of vacuum energy. However, these droplets of vacuum energy will be unstable because their mass is too small to satisfy the de Sitter conditon that a region of space with a vacuum energy willl be stable only if its radius is equal to its gravitational radius. Consequently such droplets would immediately expand and coelesce, forming a uniform vacuum energy. According to the formulae in [12] the quark exchange energy density of a uniform quark gas will be ≈1/4 the total mass-energy density for a wide range of quark densities. In our model we will simply assume that a uniform vacuum energy = ¼ of the rest mass energy density of the dust is created inside the quantum critical layer as it expands.

The absolute value of the vacuum energy will depend on the density of nucleons as they approach the quantum critical layer. In the following we will use the notation $\rho_\Lambda$ to denote the uniform vacuum energy density that is created inside the quantum critical layer as a result of the release of the quark exchange energy of the collapsing dust. In the case of the collapse of stellar cores, the nucleon density when the radius of the stellar core approaches its Schwarzchild radius will be comparable to the density of nucleons in nuclear matter, Therefore the gravitational collapse of a stellar mass dust cloud will eventually lead to vacuum energy densities $\rho_\Lambda$ similar in magniude to the vacuum energy density $B$ in the MIT "bag" model for a single nucleon [13]. After a uniform vacuum energy appears the space-time inside the quantum critical layer will resemble a Lemaitre-Robertson-Walker closed universe [14]. This interior space-time will not be stationary, but will continue to evolve due to both gravity and the radiation of nucleon decay products from the critical layer. Eventually a stationary stable compact object whose mass is almost entirley due to vacuum energy will be left behind.

Because in our theory of the formation of compact objects an event horizon never forms, all the nucleon decay products other than the vacuum energy will eventually escape. Moreover because of their weak interactions, some of the neutrinos created by nucleon decay will be able to escape even before the supply of dust cloud nucleons is exhausted. Our theory for the leakage of the nucleon decay products is predicated on the assumption that these emissions take place at the outer surface of a quantum critical layer that appears when conditions become ripe for the formation of a trapped surface in the collapsing dust cloud. According to general relativity this trapped surface will start to form at $r = 0$ at a certain proper time $\tau_0$ when a photon emitted from the center of the dust cloud reaches the surface just when the radius of the collapsing dust cloud reaches its gravitational radius $R_g$. The radius $r_c$ of the trapped surface at times $\tau > \tau_0$ follows from the equation of motion of a photon in a closed Robertson-Walker space-time:

$$\chi(\tau) = \int_{\tau_0}^{\tau} \frac{c d\tau'}{R(\tau')}, \qquad (2)$$

where $\sin\chi \equiv r_c/R$. Starting at $r_c = 0$ the trapped surface will expand until a proper time $\tau_1$ when $r_c(\tau_1) = R(\tau_1)$. A light ray emitted from the surface at this time will be trapped if $R(\tau_1) = 2GM'/c^2$, the gravitational radius of the residual dust cloud plus decay products that have not already escaped. The analytical dust cloud solution [1,2] implies that $\tau_0 = \tau_c - (1+\pi/4)^3 2R_g/3c$, where $\tau_c$ is the proper time when general relativity predicts that an initially stationary dust cloud will collapse to a singularity. The analytical dust solution also implies that for $\tau_0 < \tau < \tau_1$ the dust cloud radius $R(\tau) \approx (1+\pi/4 - \chi/2)^2 R_g$. However, because of neutrino emissions from the quantum critical surface $R(\tau)$ will deviate from the analytical dust cloud solution for $\tau > \tau_0$. We have chosen to estimate $R(\tau)$ during the period $\tau_0 < \tau < \tau_1$ by using the dust cloud analytic form for $R(\tau)$, but with a time-varying gravitational radius $R_g'$ to account for neutrino radiation:

$$R(\chi) = \left(1 + \frac{\pi}{4} - \frac{\chi}{2}\right)^2 R_g'(\chi), \qquad (3)$$

where $R'_g(\chi)$ is the gravitational radius of the dust cloud when the conformal radius of the dust cloud is $\chi$. $R'_g(\chi) < R_g$ because of the neutrino radiation.

The emission of neutrinos from dust nucleons falling onto the critical layer can be estimated in the same way we have previously used to estimate the $\gamma$-ray and positron luminosities of compact astrophysical objects due to cosmic protons falling onto these objects [7,8]. In the context of gravitational collapse the role of a cosmic proton flux is played by the inward flux of nucleons in the collapsing dust cloud. As the quantum critical surface sweeps over the dust cloud some fraction $f_v$ of nucleon mass-energy in the dust is converted into neutrinos. Taking into account both the production of neutrinos from nucleons falling onto the expanding critical layer and the escape of the neutrinos from the collapsing dust cloud, the rate of change of the neutrino energy $E_v$ contained within a surface with radius $r_c$ is:

$$\frac{dE_v}{d\chi} = f_v \sin^2 \chi R_g \frac{3c^4}{2G}\left[\cos \chi - \frac{R}{3c\tau_{esc}}\sin \chi\right], \qquad (4)$$

where $\tau_{esc}$ is the mean escape time for neutrinos. For a 10 solar mass dust cloud the quantity $R_g\, 3c^4/2G$ is on the order of $10^{54}$ erg. For the purposes of this paper we will adopt the following simple expression for the neutrino escape time:

$$\frac{c\tau_{esc}}{R_g} = 2.5\sin \chi + 8000 \frac{M_\oplus}{M}(1-\sin \chi)^2 \qquad (5)$$

The first term on the r.h.s. of Eq. 5 approximates the general relativisitic time delay calculated by Ames and Thorne [15] for a massless particle to escape from the vicinity of the Schwarzschild radius to infinity. The second term is an estimate of the classical time for neutrinos to diffuse from the quantum critical layer to the surface of the dust cloud, assuming that the neutrino transport cross-section per dust nucleon is $10^{-43}$ cm². Because of the relatively long time neutrino diffusion time, the neutrino leakage is negligible until $\tau$ is very close to $\tau_1$. For $\tau > \tau_1$

$E_\nu$ will decay exponentially with an e-folding time $2.5 R'_g/c$. In Figure 1 we show the values of $R$, $R_g'$, and $r_c$ as a function of the elapsed proper time $\Delta\tau = \tau - \tau_0$, obtained by combining Eq. 3 with numerical integration of Eqs. 2 and 4, and assuming $f_\nu = 0.5$. For $\tau > \tau_1$ our analytical expressions for $R$ and $r_c$ are no longer applicable, and $R(\tau)$ was obtained by setting $R(\tau) = R_g'$.

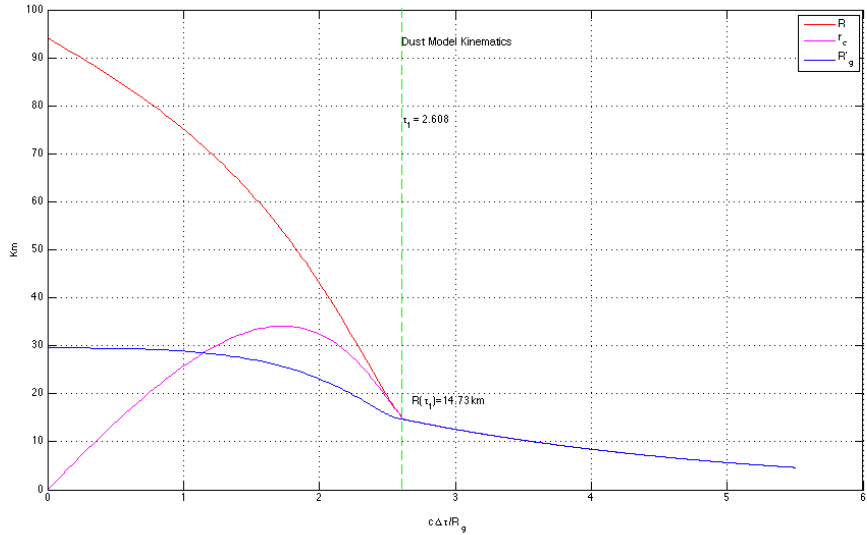

Fig.1 Radius $R$, the gravitational radius $R_g'$ of the dust cloud plus nucleon decay products, and the radius $r_c$ of the quantum critical layer for $M = 10$ solar masses as a function of the proper time interval $\Delta\tau = \tau - \tau_0$ in units of $R_g/c$. After a proper time $\tau_1$ the supply of in-falling nucleons is exhausted, and a compact object with $R \approx R_g'$ and supported by its internal vacuum energy and remains.

The emission of elementary particles other than neutrinos will be delayed because of the dust opacity; however, after the proper time $\tau_1$ there will be a large $\gamma$-ray luminosity due to photon emission from the quark-lepton plasma trapped within the quantum critical layer. It can be shown [9] that even when the effect of the radiation on the exterior space-time is taken into account, the rate of change of the mass M' of a radiating object will still be related to the surface luminosity in the familiar way:

$$\frac{dM'}{dt} = -L/c^2 , \qquad (6)$$

In this paper we will assume that the luminosity determining the mass *M'* of the collapsing dust cloud is entirely due to the radiation of neutrinos and photons from the outer surface of the critical layer. The neutrino luminosity $L_v = (\varepsilon_q/m_P c^2)^{1/4} E_v/\tau_{esc}$ seen by a distant observer as determined by Eq. 4 and assuming $f_v = 0.5$ is shown as the red line in Figure 2. The x-axis shows the time *t* in seconds as measured by an observer far from the collapsing dust cloud (but with the same cosmological red shft) that has elapsed since the nucleation of the quantum critical layer at *r* = 0. This observer time can be related to the proper time interval Δτ by matching the Robertson-Walker metric for the space-time inside the critical layer to the exterior metric. If we neglect the effect of the neutrino and γ-ray radiation on the external space-time, this matching yields:

$$t = \int_{\tau_0}^{\tau} \frac{d\tau'}{\left[1 - R'_g(\tau')/R(\tau')\right]^{1/2}} \qquad (7)$$

The time $t_1$ in Fig.2 corresponds to the proper time $\tau_1$ when $r_c = R$ and the supply of in-falling nucleons is exhausted. For proper times very close to $\tau_1$ and afterwards the observer time $t$ is related to the proper time interval $\Delta\tau$ by a constant gravitational red-shift $(m_P c^2/\varepsilon_q)^{1/4} \approx 10^5$.

As the nucleons in the dust cloud disappear, a dense plasma containing electrons, positrons, γ-rays, and, at least initially, quarks, anti-quarks, and gluons will be created inside the critical layer. For $\tau < \tau_1$ and even for some time after $\tau_1$ this plasma will be confined inside the critical layer, which implies that a very large pressure gradient will exist at the critical layer. This pressure gradient will halt at least momentarily the gravitational collapse of the nucleon decay products that are created at the critical surface. Based on the Einstein equations for a Robertson-Walker space-time with a cosmological constant [14], we would expect that the decay products inside the critical surface will then resume collapse, be stable, or expand depending on whether $\rho_m + 3p_m + \rho_v - 2\rho_\Lambda$ is positive, zero, or negative, where $\rho_m$ is the matter energy density, $p_m$ is the matter pressure, $\rho_v$ is the average energy density of the neutrinos that have not yet escaped, and $\rho_\Lambda$ is the vacuum energy density. Initially the quantity $\rho_m + 3p_m + \rho_v - 2\rho_\Lambda$ will be positive However, if our estimates that the neutrinos carry away approximately 50 percent the rest mass-energy of the nucleons and 25 percent of the nucleon mass-energy gets converted into vacuum energy are accurate, then as a result of neutrino and photon radiation the quantity $\rho_m + 3p_m + \rho_v - 2\rho_\Lambda$ will pass through zero shortly after the proper time $\tau_1$. At this point the decay products will expand to fill up the volume inside the gravitational radius $2GM'/c^2$, forming a compact object which is stable against continued gravitational collapse. This compact object will not be completely stationary though because of the continuing radiation of leptons and photons from the critical surface.

In general one would expect the leakage rate for the leptons and photons trapped inside the critical layer to depend on the density of these particles at the surface. However, in contrast to the variation in photon density inside an ordinary star where there is a steep gradient in the photon density from the center of the star to the surface, in our model the density gradient is reversed because there is a tendency to expel all the nucleon decay products from the interior. The net result is that once the opacity of the in-falling dust is removed, the γ-ray radiation rate will to a good approximation simply be $c\rho_m/2$ multiplied by the surface area, where $\rho_m$ is the average interior energy density of the plasma containing the elementary particles other than neutrinos resulting from nucleon decay. If we assume that the nucleon decay products are confined inside the quantum critical surface, we obtain the following estimate for the γ-ray luminosity seen by a distant observer for times $t > t_1$:

$$L_\gamma = \frac{3c^5}{4G}\left(\frac{\varepsilon_q}{m_P c^2}\right)^{1/4} \frac{\rho_m}{\rho_m + \rho_\Lambda + \rho_m} \tag{8}$$

Eg. 8 assumes that for $\tau > \tau_1$ the radius $R$ of the residual compact object is always equal to the gravitational radius $2GM'/c^2$. The energy densities $\rho_m$, $\rho_\Lambda$, and $\rho_\nu$ to be used in Eq. 8 can be determined self-consistently by using our previous result for the neutrino contribution to $M'$ as a function of time together with Eq. 2. The result of such a calculation is shown as the blue line in Fig. 2. This calculation assumed that at the proper time $\tau_1$ $\rho_m = \rho_\Lambda$ and the total contribution of $\rho_m$ and $\rho_\Lambda$ to $M'$

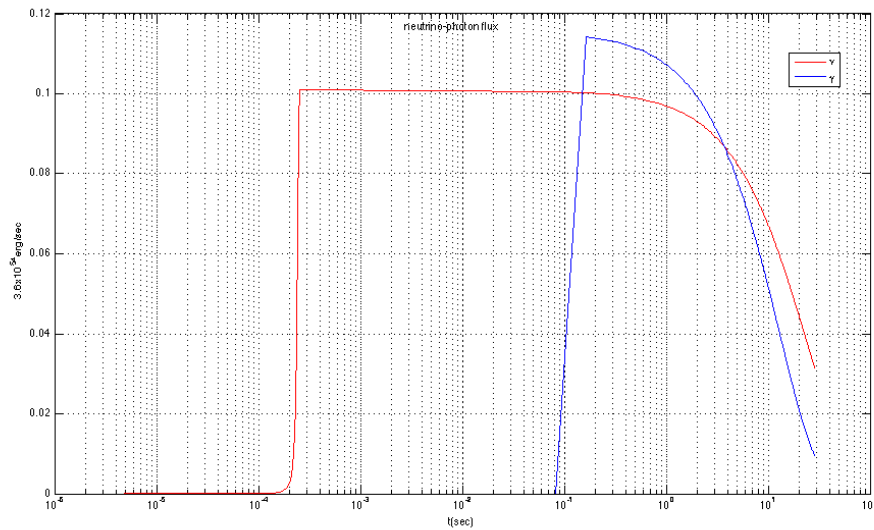

Fig. 2. The red line shows the neutrino luminosity seen by a distant observer that results from dust falling onto the quantum critical transition layer. The blue line shows the $\gamma$-ray emission seen by a distant observer coming from the plasma trapped inside the quantum critical transition layer after the time $t_1$. The x-axis shows the observer time that has elapsed since the nucleation of the quantum critical layer at $r=0$. The luminosity is measured in units of $10^{-5}c^5/G = 3.6 \times 10^{54}$ erg/sec, and it is assumed that the initial mass $M = 10$ solar mass.

= $M/2$. We have not attempted to evaluate the $\gamma$-ray luminosity for times $t < t_1$, but it will be suppressed due to the dust opacity. As is evident from Eq. 8 the peak $\gamma$-ray luminosity will be on the order of $3c^5/8G$ times the red-shift factor $(\varepsilon_q/m_Pc^2)^{1/4}$; i.e. ~ $10^{54}$ erg/sec. It is noteworthy that this peak luminosity is nearly independent of the initial mass. The spectrum of the $\gamma$-rays will be softer than that previously calculated for nucleons falling onto a compact object [7,8] due to thermalization.

Our most interesting result is that due to the gravitational red-shift $(m_Pc^2/\varepsilon_q)^{1/4}$ the predicted time duration for our neutrino and $\gamma$-ray bursts is orders of magnitude longer than the classical collapse time. This is a direct result of the prediction in Refs. 3-4 that the quantum corrections to Einstein's general relativity are singular; which in the case of gravitational collapse means that the thickness of the critical layer where classical general relativity fails is a macroscopic quantity. Indeed identification of natural bursts of neutrinos or $\gamma$-rays coming directly from the surface of a compact object with a peak luminosity and duration close to our predictions would be *prima facie* evidence for the singular breakdown of classical general relativity. As it happens the mysterious extragalactic γ-ray bursts that have been observed for some time by space-based γ-ray observatories [16] do in fact have typical luminosities and durations close the peak luminosity and duration of our bursts. If the collapsing object is surrounded by matter that doesn't collapse, then the γ-ray burst will be hidden from view; however, the neutrino burst may still be visible. These possibilities will be examined in more detail in a future paper.

In conclusion we have provided a plausible argument that in virtually all cases of gravitational collapse involving ordinary matter a residual compact object with finite size and energy density will be formed. If this residual object has a mass greater than the maximum mass for a neutron star, the residual object wiil apparently resemble either a type of compact object referred to as "dark energy star" [17] or a similar type of compact object referred to as "gravastar" [18] Recently it has been suggested [19] that as a consequence of Hawking's prediction that black holes should radiate thermal radiation a very high density of radiation should appear near the event horizon of a black hole. Athough this result appears superficially similar to our picture, it is based on an ill-fated attempt to reconcile Hawking's prediction with quantum mechanics ; and in contrast with our theory of particle emission from a quantum critical layer there are no emissions of quanta with typical energies on the order of $\varepsilon_q$. Of course, it may appear that our model begs the question as to whether black holes can in principle exist because one can imagine that the collapsing matter doesn't necessarily contain baryons. However, the replacement of a trapped surface in classical general relativity by a quantum critical transition layer ought to be a generic phenomena with typical energies of the quanta in the collapsing matter replacing $\varepsilon_q$, and the contribution of SUSY symmetry breaking to the vacuum energy replacing $\rho_\Lambda$.

### Acknowledgements
The authors are very grateful to E. Bloom, P. Joshi, A. Kerman, R. Laughlin, P. Mazur, and E. Mottola, for enlightening discussions.